\documentclass[usenatbib]{mn2e}
\usepackage{epsf}



\newcommand{\be}{\begin{equation}}
\newcommand{\ee}{\end{equation}}

\newcommand{\Msun}{M_{\odot}}

\title[Properties of Intracluster Stars]
{Simulating Galaxy Clusters - III: Properties of the Intracluster Stars}
\author[J. Sommer-Larsen, A. Romeo and L. Portinari]{Jesper
    Sommer-Larsen$^{1,2,3}$\thanks{E-mail: jslarsen@tac.dk}, Alessio D. 
    Romeo$^{1,2,4}$\thanks{E-mail: aro@ct.astro.it}
    and Laura Portinari$^{1,5}$\thanks{E-mail: lporti@utu.fi}\\
    $^1$Theoretical Astrophysics Center,
    Juliane Maries Vej 30, DK-2100 Copenhagen, Denmark\\
    $^2$Nordita, Blegdamsvej 17, DK-2100 Copenhagen, Denmark\\
    $^{3}$ Astronomical Observatory, University of Copenhagen, 
    Juliane Maries Vej 30, 2100 Copenhagen \O, Denmark\\    
    $^{4}$Dipartimento di Fisica e Astronomia, Universit\`{a} di Catania, via 
    S.Sofia 64, 95123 Catania, Italy\\
    $^5$ Tuorla Observatory, V\"ais\"al\"antie 20, FIN-21500 Piikki\"o, 
    Finland}

\begin{document}
\date{Accepted ---. Received ---; in original form 2004 March 11}

\pagerange{\pageref{firstpage}--\pageref{lastpage}} \pubyear{2004}

\maketitle

\label{firstpage}

\begin{abstract}
Cosmological ($\Lambda$CDM) TreeSPH simulations of the formation and evolution
of galaxy groups and clusters have been performed. The simulations invoke
star formation, chemical evolution with non-instantaneous recycling, 
metallicity
dependent radiative cooling, strong star-burst, and (optionally) AGN, driven 
galactic
super-winds, effects of a meta-galactic UV field and thermal conduction.
Results for two clusters, one Virgo-like
($T$$\simeq$3 keV) and one (sub) Coma-like ($T$$\simeq$6 keV), are presented.
At $z$=0 the stellar contents of both clusters consist of a central
dominant (cD) galaxy surrounded by cluster galaxies and intracluster
stars. 
The intracluster (IC) stars are found to contribute 20-40\% of the
total cluster B-band luminosity at $z$=0 and to form at a mean redshift
$\bar{z}_f\sim$3 being on average about 0.5 Gyr older than the stars in 
cluster galaxies. 
UBVRIJHK surface brightness profiles of the IC star populations are 
presented; the profile of the larger cluster matches the observed V-band
profile of the cD in Abell 1413 ($T\simeq$8 keV) quite well. 
The typical colour of the IC stellar population is B-R=1.4-1.5, 
comparable to the colour of sub-$L^*$ E and S0 galaxies.
The mean Iron abundance of the
IC stars is about solar in the central part of the cluster ($r\sim$100 kpc)
decreasing to about half solar at the virial radius. The IC stars
are $\alpha$-element enhanced with a weak trend of [O/Fe] increasing
with $r$ and an overall [O/Fe]$\sim$0.4~dex, indicative of dominant
enrichment from type II supernov\ae.

The IC stars are kinematically significantly colder than the cluster galaxies:
The velocity dispersions of the IC stars are in the inner parts of the
clusters ($r\sim$100-500 kpc) only about half of those of the cluster galaxies
increasing slightly outward to about 70\% at $r$=1-2 Mpc. 
The typical projected
velocity dispersion in the Virgo-like cluster at $R\ga$50 kpc is 300-600 km/s
depending
on orientation and projected distance from the cluster center. Rotation is
found to be dynamically insignificant for the IC stars. The velocity 
distributions of IC stars {\it and} clusters galaxies are in one cluster
highly radially anisotropic, in the other close to isotropic.
\end{abstract}


\section{Introduction}
Current hierarchical or ``bottom-up'' structure formation theories
predict that 
the field star populations of haloes of galaxies like the 
Milky Way should consist partly of stars originally born in small 
proto-galaxies and later tidally stripped from these by the main galaxy
or through interaction with other proto-galaxies. 
The fraction of halo field stars of such origin 
depends 
(apart from obvious cosmic variance
effects) on the properties of the dark matter
(e.g., cold dark matter, CDM, vs. warm dark matter, WDM: for WDM the hierarchy
is broken at the free-streaming mass scale, e.g., Sommer-Larsen \& Dolgov
2001) and the nature of the star-formation process. For example, some
halo stars may be born in rapidly expanding, supernova driven proto-galactic 
supershells and hence be detached from their parent proto-galaxies right from 
the outset in space, as well as phase-space \citep[e.g.,][]{M.97,SL.03}. 
If dark matter is cold and star-formation in expanding supershells 
is not a common phenomenon the above fraction 
could be quite large, in principle 
at least, 
approaching unity \citep[e.g.,][]{H.03}. 
The halo stars resulting from tidal stripping
or disruption of a proto-galaxy will stay localized in phase-space for
a long period and several such ``streams'' of halo stars have been
detected in the haloes of the Milky Way and M31 (e.g., Helmi et al. 1999;
Ferguson et al. 2002).

From the point of view of structure formation, galaxy clusters can be seen
as scaled up versions of galaxies in hierarchical scenarios. In particular,
gravity is expected to strip or disrupt cluster galaxies in a similar
way as satellite galaxies in galaxy haloes and a population 
of intracluster (or ``field'') stars 
should thus reside between the cluster galaxies. It has been known for
decades that cD galaxies often are embedded in extended envelopes
presumed to consist of
stars tidally stripped off galaxies in the process of
being engulfed by the cD \citep[e.g.,][]{O76,D79}. 
In recent years it has been possible to perform 
quantitative studies of these stellar envelopes through ultra-deep
surface photometry of the general stellar population 
\citep[e.g.,][]{G.00,G.04,F.02,F.04a}, or
imaging/spectroscopy of individual planetary nebulae 
\citep[e.g.,][]{A.02,A.03,F.04b} or supernovae Type Ia \citep{G.03}. The
potential importance of intracluster stars in relation to the chemical
enrichment of the intracluster medium (ICM) was recently discussed by
\cite{Z.04} and Lin \& Mohr (2004).

On the theoretical front \cite{N.03} have used an N-body dark matter only
fully cosmological simulation of the formation of a Virgo-like cluster to
make predictions about intracluster stars. They find that unrelaxed velocity
distributions and (bulk) streaming motions of the IC stars should be common
due to the large dynamical timescales in clusters. Hence the evolution
towards a relaxed coarse-grained phase-space distribution function through
phase-mixing has proceeded much less far\footnote{A visualized example from
one of our simulations is given at
http://www.tac.dk/\~~\hspace{-1.4mm}jslarsen/ICstars} 
than, e.g., in the Galactic halo, 
where dynamical timescales are much shorter. 
  
The dark matter only simulations are complemented by various N-body
simulations which invoke a more realistic modelling of the stellar properties
of galaxies in order to study the fate of galaxies in a cluster environment
(e.g., ``galaxy harassment'', Moore et al. 1996, ``tidal stirring'', Mayer
et al. 2001 and the formation of the central cluster galaxy, Dubinski 1998).
The advantage of this approach is that a very high resolution in the
stellar component can be obtained. In relation to the properties of the IC
stars, recent progress has been made by \cite{D.03}, and \cite{F.04a}.

In general the properties of the system of IC stars are set by two main
effects: a) the cool-out of gas and subsequent conversion of cold, 
high-density gas to stars in individual galaxies and b) the 
stripping/disruption of the galaxies through interactions with other
galaxies and the main cluster potential. Since such interactions will
generally affect the star-formation rate (as long as a reservoir of gas
is available) the former process is intimately coupled to the latter.

Only fairly recently has it been possible to carry out fully cosmological 
gas-dynamical/N-body simulations of the formation and evolution of galaxy
clusters at a sufficient level of numerical resolution and physical 
sophistication that 
the cool-out of gas, star-formation,
chemical evolution and gas inflows and outflows related to
individual cluster galaxies can be modelled to, at least some, degree of 
realism \citep[e.g.,][]{V03,T.04}. Recently \cite{M.04} analyzed 117
clusters formed in a large cosmological TreeSPH simulation \citep{B.04} to
study the properties of the IC stars. They determined average density
profiles of the IC stars as well as stars in galaxies for their large sample
of clusters. One of the interesting conclusions reached is that only at
fairly large cluster-centric distances ($r\ga$0.4-0.5$r_{\rm{vir}}$) does the
(spherically averaged) density of stars in galaxies (excluding the cD) exceed 
that of IC stars. Complementary work has recently been performed by 
\cite{W.04}, who simulated a Coma-like cluster at very high numerical 
resolution. They find an IC star fraction of 10-20\%, in good agreement
with observations, and a tendency for the fraction to increase with time.
In relation to on-going observational efforts, they moreover find that 
line-of-sight velocities of intra-cluster planetary 
nebulae can be used for fairly accurate cluster mass estimation, even for
dynamically quite unrelaxed clusters, if more than 5 fields at a range
of projected radii are covered.

We have recently undertaken similar
simulations of galaxy groups and clusters. We are building on our
TreeSPH code used for simulating galaxy formation \citep[e.g.,][]{SL.03},
improved to include modelling of non-instantaneous chemical evolution 
\citep{L.02}, metallicity-dependent, atomic radiative cooling, strong 
supernova, and (optionally) AGN, driven galactic winds and thermal 
conduction. 
The main results of the simulations concerning the properties
of the ICM and cluster scaling relations are described in Paper~I 
\citep{R.04}, while the characteristics of the population of the simulated
cluster galaxies are presented in Paper~II \citep{RPSL04} ---
in this Paper~III of the series, we discuss our first results on the
properties of the IC stars in two simulated clusters, with masses and 
temperatures similar to the Virgo and Coma clusters, respectively.
Our results complement those of \cite{M.04} and \cite{W.04}, since we present 
for the IC stars also results on self-consistently calculated multiband 
surface brightness profiles, 
colours, abundances and detailed kinematics (Willman et al. 2004 also present
a kinematics study, which we compare to in this paper).

In section 2 we briefly describe the code and the numerical simulations, 
in section
3 we present and discuss the results obtained, and finally section 4 
constitutes our conclusions.
\section{The code and simulations}
The code used for the simulations is a significantly improved version of
the TreeSPH code we have used previously for galaxy formation simulations 
\citep{SL.03}: Full details on the code are given in Paper~I, here we
recall the main improvements over the previous version.
(1) In lower resolution regions (which will always be
present in cosmological CDM simulations) an improvement in the numerical
accuracy of the integration of the basic equations is obtained by
solving the entropy equation rather than the thermal energy equation ---
we have adopted the ``conservative'' entropy 
equation solving scheme suggested
by \cite{SH02}. 
(2) Cold high-density gas is turned into stars in a probabilistic way as
described in \cite{SL.03}. In a star-formation event an SPH particle
is converted fully into a star particle. Non-instantaneous recycling of
gas and heavy elements is described through probabilistic ``decay'' of star 
particles back to SPH particles as discussed by \cite{L.02}. 
In a decay event a star particle is converted fully into an SPH particle.
(3) Non-instantaneous chemical evolution tracing
10 elements (H, He, C, N, O, Mg, Si, S, Ca and Fe) has been incorporated
in the code following Lia et~al.\ (2002a,b); the algorithm includes 
supernov\ae\ of type II and type Ia, and mass loss from stars of all masses.
Metal diffusion in Lia et al.\ was 
included with a diffusion coefficient $\kappa$ derived from models of the 
expansion of individual supernova 
remnants. A much more important effect in the present simulations, however,
is the redistribution of metals (and gas) by means of star-burst driven
``galactic super--winds'' (see point 5). This is handled self-consistently by
the code, so we set $\kappa$=0 in the present simulations. 
(4) Atomic radiative cooling depending both on the metal abundance
of the gas and on the meta--galactic UV field, modelled after Haardt
\& Madau (1996) is invoked. We also include a simplified treatment
of radiative transfer, by switching off the UV field where the gas
becomes optically think to Lyman limit photons on scales of $\sim$ 1~kpc.
(5) 
Star-burst driven, galactic super-winds are incorporated in the simulations.
This is required to expel metals from the galaxies and get the 
abundance of the ICM to the observed level of about 1/3 solar in iron. 
A burst of star formation is modelled in the same way as the ``early 
bursts'' of \cite{SL.03}, i.e.\ by halting cooling in the surrounding gas 
particles, to mimic the initial heating and subsequent adiabatic expansion 
phase of the supershell powered
by the star-burst; this scheme ensures effective energy
coupling and feedback between the bursting star particle and the surrounding 
gas. The strength of the super-winds is modelled
through a free parameter $f_{\rm{wind}}$ which determines how large a fraction
of the new--born stars partake in such bursting, super-wind 
driving star formation. We find that
in order to get an iron abundance in the ICM comparable to observations,
$f_{\rm{wind}}\ga0.5$ and at the same time a fairly top-heavy Initial Mass
Function (IMF) has to be used. 
(6) 
Thermal conduction was implemented in the code following \cite{CM99}.  

We ran simulations of two small and two larger galaxy groups as well as a 
Virgo-like and a (mini) Coma-like cluster --- we shall denote the latter two
clusters C1 and C2 in the following. All 6 systems were chosen to be 
fairly relaxed (no $\ga$1:2 merging at $z\la$1). Virial masses at $z$=0 
were approximately 3.4x10$^{13}$, 7.9x10$^{13}$, 2.0x10$^{14}$ and 
8.4x10$^{14}$
$h^{-1}$M$_{\odot}$ and X-ray 
emission weighted temperatures 1.1, 1.6, 3.0
and 6.0 keV, respectively. The groups and clusters were selected from a
cosmological, DM-only simulation of a flat $\Lambda$CDM model, with
$\Omega_M$=0.3, $\Omega_b$=0.036, $h$=0.7 and $\sigma_8$=0.9 and a box-length
of 150 $h^{-1}$Mpc. Mass and force resolution was increased in Lagrangian
regions enclosing the groups and clusters, such that $m_{\rm{gas}}$=$m_*$=
2.5x10$^8$ and $m_{\rm{DM}}$=1.8x10$^9$ $h^{-1}$M$_{\odot}$ for the high 
resolution gas, star and dark matter particles. Particle numbers are in the
range 100000-500000 
baryonic+DM particles. Gravitational (spline) 
softening lengths of $\epsilon_{\rm{gas}}$=$\epsilon_*$=2.8 and 
$\epsilon_{\rm{DM}}$=5.4 $h^{-1}$kpc, respectively, were adopted. The
gravity softening lengths were fixed in physical coordinates from $z$=6
to $z$=0 and in comoving coordinates at earlier times.
To test for numerical resolution effects one simulation was run with 
eight times higher mass resolution and two times higher force
resolution, yielding 
$m_{\rm{gas}}$=$m_*$=
3.1x10$^7$ and $m_{\rm{DM}}$=2.3x10$^8$ $h^{-1}$M$_{\odot}$ and gravity
softening lengths of 1.4
and 2.7 $h^{-1}$kpc, respectively. 


For the simulation presented in this paper $f_{\rm{wind}}$=0.8,
and an Arimoto-Yoshii IMF (of slope $x$=--1, shallower than the 
Salpeter slope $x$=-1.35) with mass limits [0.1--100]~$\Msun$
was adopted. AGN driven winds were not invoked.
Moreover, the thermal 
conductivity was set to zero assuming that thermal conduction in the ICM is 
highly suppressed by magnetic fields \citep[e.g.,][]{EF00};
notice however that, in relation to the cluster galaxy populations 
and hence also IC stars, no significant difference is found between
simulations with zero thermal conductivity and simulations invoking thermal
conduction at 1/3 of the Spitzer level (Paper~II).
\section{Results and Discussion}
In this section we present and discuss results for clusters C1 and C2
run at our ``standard'' resolution. The resolution test discussed in 
Section 3.7 indicates that none of the following results concerning
mass and luminosity fractions, mean ages, surface brightness profiles,
colours, abundances, kinematics and density profiles of the IC stars
change in any significant way with increased numerical resolution.

All results presented in the following refer to $z$=0, unless another epoch
is explicitly specified.
At $z$=0 1.2x10$^4$ and 2.3x10$^4$ star particles are located inside of
$r_{\rm{vir}}$=1.3 and 2.0~$h^{-1}$Mpc of the clusters C1 and 
C2\footnote{In this paper the virial radius
$r_{\rm{vir}}$ is the radius corresponding to, at $z$=0, an over-density 337 
times the
mean density of the Universe, as appropriate for a top--hat collapse
in the adopted $\Lambda$CDM cosmology (e.g., Bryan \& Norman 1998). The 
virial mass
$M_{\rm{vir}}$ of the cluster is the mass enclosed within the region of radius 
$r_{\rm{vir}}$.}. The 
corresponding mass in stars is 2.9x10$^{12}$ and 5.6x10$^{12}$ 
$h^{-1}$M$_{\odot}$ or a fraction of about 1\% of the total virial mass
(the stellar mass fraction is smaller for C2 than for C1 - we shall return
to this below). In both clusters the stellar mass constitutes 14\% of the
total baryonic mass inside of the virial radius. In the following we shall
refer to star particles simply as ``stars''.

When preparing 
the initial conditions for C2, due to computational
limitations we sampled with split DM+SPH particles a region somewhat
smaller than the``virial volume''. The volume outside this region, but still
inside of the virial volume, is
represented by high resolution, DM-only particles (with a mass scaled by
$1/(1-f_b)$ with respect to the DM particles in the inner DM+SPH region, 
since they are not split into baryonic and DM component; $f_b$ is the 
universal baryon fraction, assumed to be 0.12 in this work); finally, the 
region outside the virial volume is, as usual, sampled by lower resolution 
dark matter particles with masses increasing with distance from the virial 
volume.
As a consequence of this baryonic under-sampling, about half of the mass 
inside of the virial radius of the final cluster is in the form of 
(high--resolution) DM-only particles, i.e.\ not split into SPH+DM. 
The relative percentage of these DM-only particles increases with radius,
hence the amount of late infall of gas and galaxies is reduced compared
to the fully resolved case. This explains why the stellar mass fraction
in C2 is smaller than that in C1 as mentioned above (the C2 star mass
is only about twice the C1 star mass, while the respective virial masses differ
by a factor of about four). The ratio of the respective number of galaxies 
$N_{\rm{gal}}$(C2)/$N_{\rm{gal}}$(C1) is also smaller than the ratio of the
corresponding virial masses (again, a factor of $\sim$2 vs.\ a factor of 
$\sim$4, 
see below) and the average ages of the IC stars and galaxies in C2 are 
slightly larger
than that in C1, due to the under-sampling of late infall of galaxies
at the outskirts (Section 3.3). Since the results we discuss refer 
to the inner 1--2~Mpc, the baryonic under-sampling of the outskirts is not
a crucial problem.
Moreover, we have initiated a simulation of C2 with full baryonic sampling 
of the virial volume;
at a redshift of $z=0.8$ this shows no significant difference in relation to 
the results on IC stars presented here.

The stellar contents of both clusters are characterized by a massive, central
dominant (cD) galaxy surrounded by galaxies and intracluster stars. The
effective radii, calculated from the B-band surface brightness profiles of the
two cD galaxies, are $R_{eff}\simeq$5 kpc, taking the inner 80 kpc of the 
cluster as the extent of the cDs --- see below. After correction for an excess
of central, young stars (section 3.2) the absolute magnitudes of the cDs are
$M_B \sim$ -23. Extrapolating the observational ``Kormendy relation'' 
\citep{K77} such bright cDs have $R_{eff}\sim$ 20$h^{-1}$ kpc. Part of this
discrepancy is due to an excess of central, young stars, as we shall
discuss in section 3.2. 
We also note that in a study of the formation of the brightest 
cluster galaxy (BCG) based
on a pure N-body simulation (but of higher resolution than ours) \cite{D98} 
found an effective radius of $\sim$20 kpc for the central BCG.
\subsection{Identifying Cluster Galaxies and Intracluster Stars}
The first step in identifying the IC stars is to find all galaxies in
the clusters. To this end we proceed in the following simple way: by visual
inspection of the $z$=0 frames the stars in all galaxies except the cD are
located within 10-15 kpc from the centers of the galaxies. Guided by this
typical size of the individual cluster galaxies, a cubic grid of
cube-length $\Delta l$=10 kpc is overlaid the cluster, and all
cubes containing at least $N_{\rm{th}}$=2 stars are identified. Subsequently,
each selected cube is embedded in a larger cube of cube-length 3$\Delta l$.
If this larger cube contains at least $N_{\rm{min}}$=7 star particles, which 
are 
gravitationally bound by its content of gas, stars and dark matter the system
is identified as a potential galaxy. Since the method can return several,
almost identical versions of the same galaxy only the one
containing the largest number of star particles is kept and classified
as a galaxy. We tested the galaxy identification algorithm by varying
$\Delta l$, $N_{\rm{th}}$, $N_{\rm{min}}$, and also the numerical resolution
of the simulations (section 3.7) and found it to be adequately robust for the
purposes of this paper. 

The algorithm identifies $N_{\rm{gal}}$=42 and 94 galaxies
inside of the virial radii of C1 and C2, respectively, plus a cD at the 
center of each cluster. 
The ratio $N_{\rm{gal}}$(C2)/$N_{\rm{gal}}$(C1)
is smaller than the ratio of the corresponding virial masses due to the 
baryonic under-sampling in cluster C2, as discussed above.
Apart from the
cD there are no galaxies inside of $r_{\rm{cD}}$=80 kpc in either of the $z$=0 
frames (this is not true in all $z\sim$0 frames for clusters C1 and C2, 
though, but we tested that the results we present in the following did not
depend on which $z\sim$0 frame was chosen to represent the present epoch ---
see also below).

We define the system of IC stars as the stars located at cluster-centric
distances $r_{\rm{cD}}\le r \le r_{\rm{vir}}$ and {\it not} inside of
the tidal radius of any galaxy in the cluster. The tidal radius 
for each galaxy is taken to be the Jacobi limit
\begin{equation}
r_J = \left(\frac{m}{3M}\right)^{1/3}\hspace{-1mm}D ~~~,
\end{equation}
where $m$ is the mass of stars, gas and dark matter in the galaxy (inside of
$r_J$), $D$ is the distance from the cluster center to the galaxy and 
$M$ is the total mass of the cluster inside of $D$ 
\citep[e.g.,][]{BT87}.  The cD
itself is effectively just the inward continuation of the system of IC
stars and the division between IC stars and cD is somewhat arbitrary
(we hence below quote intracluster star fractions for $r_{\rm{cD}}$=80 as well
as 40 kpc). 
The above definition of IC stars is conservative, since
we calculate the tidal radii on the basis of the $z$=0 frame 
cluster-centric distances of the galaxies. For any 
galaxy which has been through at least one peri-center passage,
this tidal radius should be taken as a firm upper limit.
Moreover, some IC stars will be inside of the tidal
radius of one of the cluster galaxies just by chance. As the total ``tidal
volume'' of all the cluster galaxies is found to be a few tenth of a 
percent of the virial volume of the cluster this effect can be neglected,
however.

\begin{figure}
\epsfxsize=\columnwidth
\epsfbox{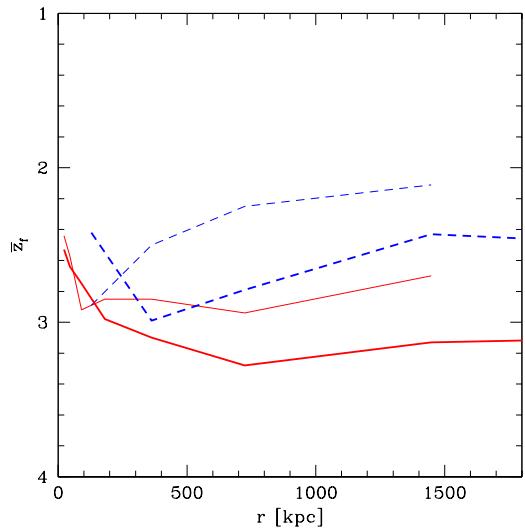}
\caption{Mean redshift of formation of the cD + IC stars (solid lines) and
stars in galaxies (dashed lines) at $z$=0. Results for cluster C1 
(``Virgo-like'') are shown by thin lines and for C2 (``mini-Coma'') by
heavy lines. Statistical uncertainties are $\Delta\bar{z}_f\la$0.05.}
\label{fig1}
\end{figure}
\subsection{The Intracluster Star Fraction}
Using the above definition we find that the intracluster stars
constitute 21 and 28\% of the stellar mass
inside of the virial radius for clusters C1 and C2,
respectively (adopting $r_{\rm{cD}}$=40 kpc these numbers increase 
slightly to 28 and 34\%). For comparison with observations it is more 
relevant to look at stellar luminosities. Since ages and metallicities are
available for all star particles, the photometric properties are 
straightforward to calculate treating each star particle as a Single
Stellar Population (SSP; see Paper~II for details). 
SSP luminosities are computed 
by mass-weighted integration of the Padova isochrones \citep{G.02}, 
according to the Arimoto-Yoshii IMF. We find that 
the IC stars contribute only 9 and 11\% of the cluster B-band luminosities
for C1 and C2 respectively. This is significantly less than the
observational estimate of $L_{\rm{B,IC}}/L_{\rm{B,tot}}\ga$0.2 
\citep[e.g.,][]{A04}. The likely reason for this discrepancy is an excess
of stars formed fairly recently at the center of the cD and dominating
the B-band luminosity: 
After a period of major merging at $z\sim$1-2 strong,
quasi-stationary cooling flows develop at the centers of the clusters despite 
the strong, super-nova driven energy feedback to
the IGM/ICM through galactic super-winds and the use of a fairly top-heavy
IMF. As a consequence, stars continue to form steadily at the centers of the
cDs ($r\la$10 kpc) at rates of $\sim$350 and $\sim$600 $M_{\odot}$/yr at $z$=0
for C1 and C2, respectively.
 At $z$=0 the cDs are too blue compared to observed cDs
with central B-R colours of $\sim$1.0 (see Fig.~4) rather than 1.4-1.5 
(note though that some
star-formation is observed in many cDs at the base of the cooling flow ---
e.g., McNamara 2004). To make a (albeit crude) correction for this excess
of young stars at the centers of the simulated cDs we temporarily discard 
the luminosity contribution of all stars in the inner $r_c$=10 kpc with 
formation redshifts $z_f<z_c$,
where $z_c$ is determined by requiring that the mean formation redshift
of the remaining stars inside of $r_c$ should be similar to the mean
formation redshift of stars at $r\ga r_c$. We find that this is the case for 
$z_c \sim$1 and shall in the following in various cases present results 
with and without this crude correction, adopting $z_c$=1 for the former. 
With this correction $L_{\rm{B,IC}}/L_{\rm{B,tot}}$=0.29 and 0.42 for
the two clusters in better agreement with observations (for 
$r_{\rm{cD}}$=40 kpc we obtain $L_{\rm{B,IC}}/L_{\rm{B,tot}}$=0.42 and 0.50, 
respectively), though somewhat high compared to the values found by
\cite{F.04a} for non-cD clusters. Moreover, the
effective radii of the cDs increase to 9 and 11 kpc, respectively. These values
are in better agreement with observations \citep{K77} and 
the finding of \cite{D98}, but still somewhat smaller. Part of this remaining
discrepancy may be due to the effect of gravitational ``pinching'' of our
cDs by the excess, young central stars.

Regarding the intracluster star fraction we show in Paper II 
that clusters C1 and C2 are deficient in bright galaxies apart from the cDs.
As discussed in Paper II, this is probably related to the fact that C1 and
C2 are fairly relaxed systems, for which the last major merger took place
at $z\ga$1. Hence it is possible that the galaxies in these clusters suffered
more tidal stripping and merging than in an average cluster.
To be conservative the above IC star fractions should then be 
considered upper limits to what would be found for a statistically 
representative sample of cluster merging histories. Finally, we note
that the observed IC star fraction of $\sim$20\% has also been reproduced
in the simulations of \cite{M.04} and \cite{W.04}. 
\subsection{Mean Formation Redshifts, Surface Brightness Profiles and Colours
of the Intracluster Stars} 
In Figure 1 we show the mean (spherically averaged) redshift
of formation, $\bar{z}_f$, of the cD + IC stars (solid lines) and stars
in galaxies except the cD (dashed lines) as a function of radial distance 
from the
center of the cD for clusters C1 and C2, respectively. For both clusters
the average formation redshift of the IC stars is $\bar{z}_{f,IC}\sim$3 at
$r\ga$100-200 kpc. The stars in galaxies (except the cD) are on average
somewhat younger with $\bar{z}_{f,gal}\sim$2.5. This is to be expected, since
the bulk of the IC stars originate in (proto) galaxies, which have been
partly or fully disrupted through tidal stripping in the main cluster
potential or by galaxy-galaxy interactions. In contrast, the galaxies
still remaining at $z$=0 have potentially been able to continue forming stars 
out of remaining cold gas or gas recycled
by evolved stars and subsequently cooled to star-forming temperatures
and densities. Still, due to ram-pressure stripping 
of the hot and dilute gas reservoir in galactic haloes 
and other effects, the
star-formation rate in the galaxies decreases significantly from $z$=2 to 0,
considerably more so than in field galaxies cf. Paper II. 
\cite{M.04} find a similar trend that the IC stars are on average older
than the stars still in cluster galaxies, but they find somewhat lower
mean redshifts of formation, $\bar{z}_f$=1.9 and 1.7 respectively. The reason
for this discrepancy is not clear: it may be related to different galactic
super-wind prescriptions or that we incorporate metal-dependent radiative
cooling in the simulations, whereas \cite{B.04} use a primordial cooling
function, but is probably not due to numerical resolution differences
(Murante 2004, private communication).

\begin{figure}
\epsfxsize=\columnwidth
\epsfbox{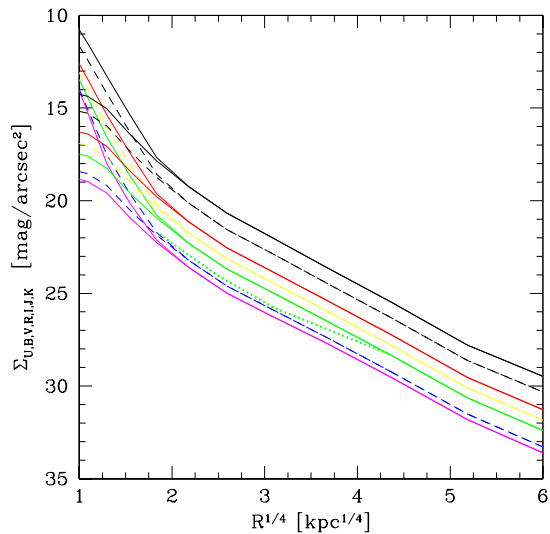}
\caption{Multiband (UBVRIJK going bottom up) surface brightness profiles of 
cD + IC stars for cluster C2, shown with (thin lines) and without (thicker
lines) the correction for the excess young, central stars, discussed in the
text. Also shown (thick dotted line) is V-band surface photometry for the
cD in the rich cluster A1413 ($T\simeq$8 keV) obtained by Feldmeier et al.
2002}
\label{fig2}
\end{figure}

\begin{figure}
\epsfxsize=\columnwidth
\epsfbox{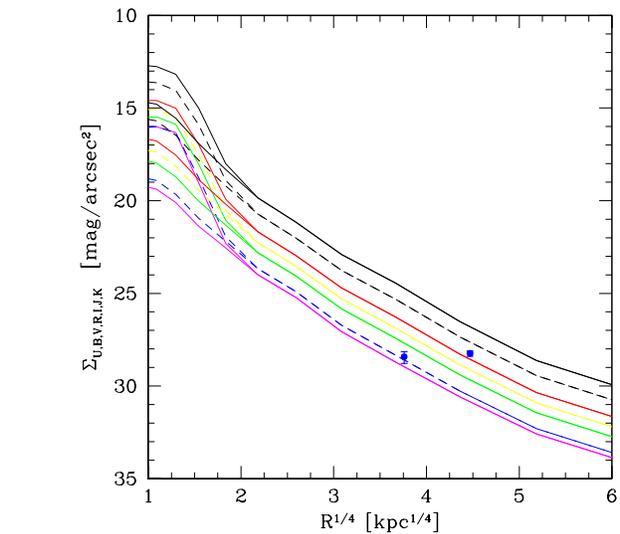}
\caption{Same as in Figure 2, but now for cluster C1. Data points are
B-band IC star estimates based on planetary nebulae observations in
the Virgo cluster by Arnaboldi et al. 2002}
\label{fig3}
\end{figure}

From Figure 1 it can be seen that 
both IC stars and galactic stars 
in cluster C2 are on average somewhat older than those in C1. 
The reason for this is numerical, rather than
physical, related to the under-sampling of baryons in the outskirts of the
virial volume of the cluster discussed at the beginning of this section:
the late infall of gas and galaxies onto the cluster is reduced with respect
to a simulation with full baryonic sampling of also the outer parts of the
virial volume. 


\begin{figure}
\epsfxsize=\columnwidth
\epsfbox{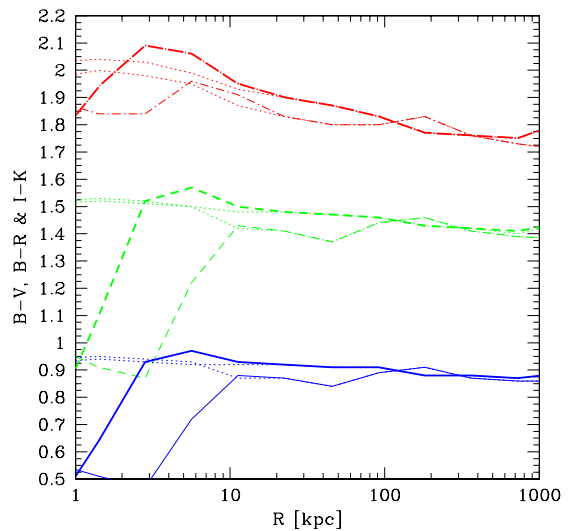}
\caption{Azimuthally averaged B-V (solid lines), B-R (dashed lines) and I-K 
(dot-dashed lines) colours of cD + IC stars for clusters C2 (thick lines)
and C1 (thinner lines), respectively. The results of applying the correction
for the excess, central young stars is shown with thin dotted lines.}
\label{fig4}
\end{figure}

In Figure 2 we show for cluster C2 the azimuthally averaged UBVRIJK
surface brightness profiles of the cD+IC stars, both with and
without the correction for central, young cD stars discussed above
(the H-band results are omitted for clarity). 
The orientation is such that the
cD is viewed along the minor axis, but other orientations give very similar
results. The light profiles are approximately
described by $r^{1/4}$ laws. In reality the slope flattens with increasing
$r$, so the system can be seen as a central dominant elliptical galaxy
surrounded by an extended envelope, 
with the slope of the surface brightness profile
becoming close to constant beyond $r\sim$40 kpc. 
These are characteristics of observed cD galaxies
\citep[e.g.,][]{F.02}. Observationally, cD+IC stars can be traced by
surface photometry to V$_{lim}\simeq$28.3 mag/arcsec$^2$ 
\citep[e.g.,][]{F.02}, which for
cluster C2 corresponds to $r\sim$350 kpc. In Figure 2 we also show V band
surface photometry of the cD + envelope in the rich cluster A1413 
($T\simeq$8 keV) obtained by \cite{F.02}. Though the observational V band 
profile is slightly flatter than what we predict for the V band, overall 
agreement is quite good, especially when it is taken into account that
the sampling of the baryons becomes increasingly incomplete with radius
in cluster C2 (see above). In Figure 3 we show the surface brightness profiles
of C1 (``Virgo''). Also shown is two B-band data points derived for Virgo IC 
stars (indirectly) using planetary nebulae by \cite{A.02}. Our predicted
B-band profile fits the inner point well, but not the outer one, which is too
high (in fact slightly higher than the inner one). The Virgo
cluster is known, however, to have an unusually flat galaxy surface brightness
profile and possibly be dynamically unrelaxed \citep{B.87}.

Shown in Figure 4 are the azimuthally averaged B-V, B-R and I-K colours
of the cD+IC stars, again with and without the correction discussed 
above. At about 10 kpc the colours (e.g., $B-V\sim$0.90-0.95) are 
typical of the stellar populations in the inner parts of cDs 
\citep[e.g.,][]{M92}. Within
10 kpc this is also the case when 
correcting for the excess of central, young stars 
--- if not, the core of the cD gets too blue
with central colours approaching those of spiral galaxies. Outside of 10
kpc the colour of the cD+IC stars gets bluer with increasing $R$, but
the gradient is very shallow: $\Delta$(B-V) per kpc of $\sim$~-0.0003
As the
mean age of the stars is approximately constant with $R$ (see Fig.~1), this
is mainly due to a decrease in metal abundance with $R$ (see below). At
100 kpc $B-R \simeq$1.45, typical of sub-$L^*$ E and S0 galaxies 
\citep[e.g.,][]{G.98}. We shall discuss this in a forthcoming paper, but note
that cDs are found in general to have quite flat colour profiles, in some
cases getting redder with $R$ \citep[e.g.,][]{M92,G.97,G.00}.  
\subsection{Abundance Properties of the Intracluster Stars and Cluster
Galaxies} 
\begin{figure}
\epsfxsize=\columnwidth
\epsfbox{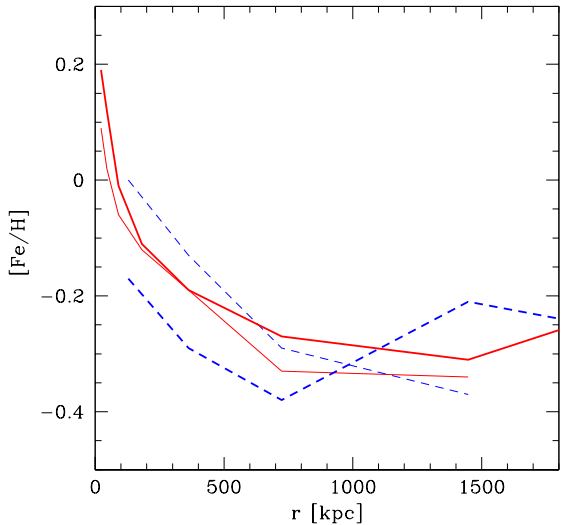}
\caption{Spherically averaged iron abundance of cD+IC stars (solid lines)
and stars in cluster galaxies (dashed lines) for clusters 
C1 (thin lines) and C2 (thick lines), respectively (results for $r<$20 kpc are
not shown for clarity).}
\label{fig5}
\end{figure}
\begin{figure}
\epsfxsize=\columnwidth
\epsfbox{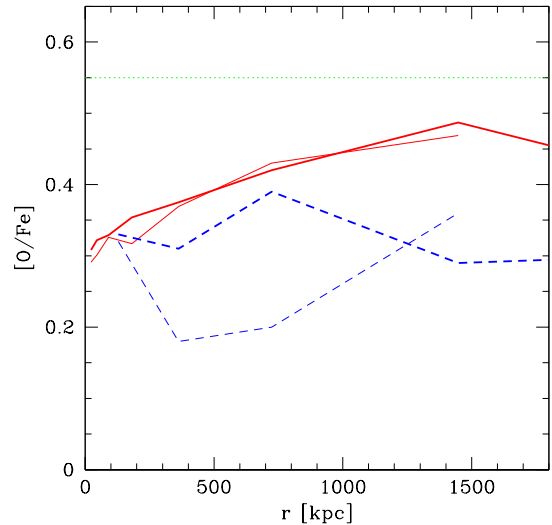}
\caption{Spherically averaged oxygen-to-iron abundance ratio of cD+IC stars 
(solid lines) and stars in cluster galaxies (dashed lines) for 
clusters C1 (thin lines) and C2 (thick lines), respectively. The limiting
case for pure SNII enrichment [O/Fe]=0.55 is shown by the thin, dotted line
(results for $r<$20 kpc are not shown for clarity).}
\label{fig6}
\end{figure}
In Figure 5 we show the spherically averaged iron abundance of the 
cD+IC stars as well as of the stars in cluster galaxies as a function
of cluster-centric distance. 
Iron is super-solar in the cD+IC stars at
$r\la$100 kpc and decreases to about half solar at large $r$; 
the stars in cluster galaxies follow a similar trend.
The fairly large overall iron abundance of the IC stars, as compared to,
e.g., stars in the halo of the Milky Way, reflects that it is the 
galaxies (past and present) which have to enrich the 5-10 times more massive
hot intracluster medium (ICM) to an iron abundance of about 1/3 solar
(Paper I). 
\cite{D.02} carried out HST observations of an IC field in the
Virgo cluster at an average projected distance of 150 kpc from M87 
(which for the purposes here can be assumed coincident with cluster center). 
They confirm an excess of red number counts, 
which they interpret as IC RGB stars. By comparison with observations 
of a dwarf irregular, they conclude that these stars have 
-0.8$<$[Fe/H]$<$-0.2. Though this is somewhat less than we predict
at $r\sim$150 kpc, the
discrepancy is only a factor of about two, and the observational abundances
are clearly significantly larger than those of the halo stars of the
Milky Way.

Figure 6 shows the corresponding oxygen-to-iron ratios as a function of $r$.
[O/Fe] is super-solar everywhere. This is in agreement with present estimates
for the luminous elliptical galaxies that contain most of the stellar mass in 
cluster galaxies. For the IC stars no observational information is currently
available (Arnaboldi 2004, private communication).

For pure type II supernovae enrichment and with the Arimoto-Yoshii IMF, 
one expects
[O/Fe]$_{\rm{SNII}}$=0.55 \citep[e.g.,][]{L.02}, so it follows
from Fig.~6 that SNe Ia do contribute somewhat to the enrichment
of the cD+IC stellar populations, and (not surprisingly) even more so for the
stars still in galaxies at $z$=0 (in fact for an Arimoto-Yoshii IMF the
global ($t\rightarrow\infty$) value of the SNII+SNIa enrichment is 
[O/Fe]=0.18).
\subsection{Kinematics of the Intracluster Stars and Cluster Galaxies}
\begin{figure}
\epsfxsize=\columnwidth
\epsfbox{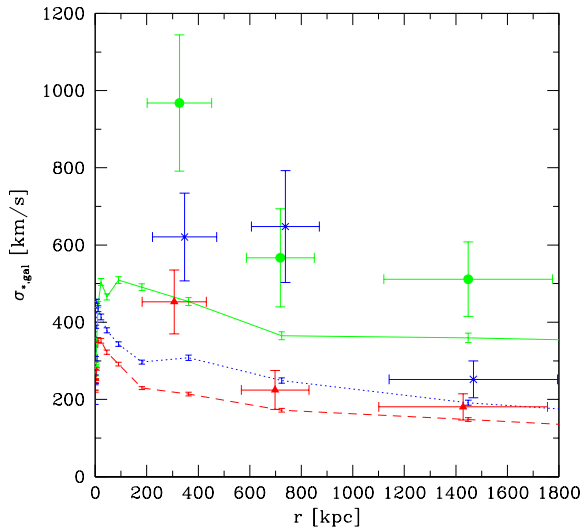}
\caption{For cD+IC stars in cluster C1 we show the
velocity dispersions
$\sigma_r$ (solid lines), $\sigma_{\phi}$ (dotted line) and 
$\sigma_{\theta}$ (dashed line). The statistical uncertainties are marked
with small errorbars. The similar quantities are shown for the cluster
galaxies by filled circles, crosses and filled triangles, with larger
errorbars marking the statistical uncertainties.}  
\label{fig7}
\end{figure}
Using observed velocities of planetary nebulae it will ultimately be 
possible to kinematically ``dissect'' the systems of cD+IC stars in nearby 
galaxy clusters, such as Virgo. It is hence of considerable interest to
determine for our simulations the velocity distribution of the cD+IC stars
and compare it to that of cluster galaxies. To this end we proceed as
follows: At $z$=0 cluster C1 and the cD at the center are somewhat
flattened (ellipticity $\epsilon=1-b/a\simeq$0.4)
with similar minor axis orientations (we shall denote the minor
axis of the cD the ``z-axis'' in the following).
Cluster C2 is only slightly flattened ($\epsilon\simeq$0.2),
and so is the cD at the center, with
approximately perpendicular minor axis orientations. For each cD+IC star
and each cluster galaxy we determine three perpendicular velocity components:
The radial component $v_r$=$\vec{v}\cdot\vec{e}_r$, where $\vec{e}_r$ is the
unit vector pointing radially away from the center of the cluster, the
perpendicular (tangential) component $v_{\phi}$=$\vec{v}\cdot\vec{e}_{\phi}$,
where $\vec{e}_{\phi}$ is the unit vector perpendicular to $\vec{e}_r$ and
aligned with the x-y plane and the third (tangential) component 
$v_{\theta}$=$\vec{v}\cdot\vec{e}_{\theta}$, where $\vec{e}_{\theta}$ is
the unit vector $\vec{e}_{\theta}=\vec{e}_r\times\vec{e}_{\phi}$.

\begin{figure}
\epsfxsize=\columnwidth
\epsfbox{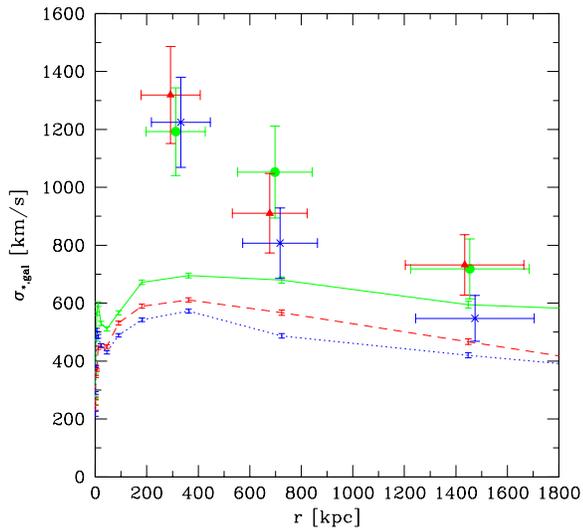}
\caption{Same as Figure 7, but for cluster C2.}  
\label{fig8}
\end{figure}
We calculate the mean rotation $\bar{v}_{\phi}$ and velocity dispersions
$\sigma_r$, $\sigma_{\phi}$ and $\sigma_{\theta}$ of cD+IC stars and
galaxies in spherical shells. Outside $r\sim$10 kpc rotation is found
to be dynamically insignificant with $\bar{v}_{\phi}\la$20-40 km/s in both
clusters. When correcting for the excess young, central stars rotation is
found to be also dynamically unimportant at $r<$10 kpc.
In Figures 7 and 8 we show the
velocity dispersions of the cD+IC stars and cluster galaxies in clusters
C1 and C2, respectively, versus cluster-centric distance. The cD+IC stars
are kinematically significantly colder than the cluster galaxies:
The velocity
dispersions of the cD+IC stars are in the inner parts of the clusters 
($r\sim$100-500 kpc) only about half of those of the cluster galaxies,
increasing slightly to about 70\% at $r\sim$1-2 Mpc. As
stars and galaxies are moving in the same gravitational potential this
implies by Jean's theorem \citep[e.g.,][]{BT87} that the (number) density
distribution of cluster galaxies is significantly flatter than that of
cD+IC stars, in particular in the inner parts of the clusters --- see below.
\cite{W.04} also find that the IC stars are kinematically colder than the
cluster galaxies, except for the innermost part of the cluster. The difference
here is probably due to the different definitions of IC stars used:
\cite{W.04} require the IC stars to be unbound from the ``cD'' also --- this
results in large IC star velocity dispersions in the inner parts of the
cluster.
\begin{figure}
\epsfxsize=\columnwidth
\epsfbox{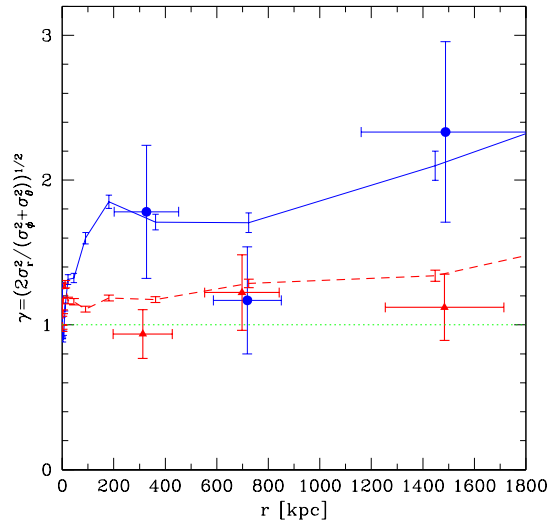}
\caption{Velocity anisotropy parameter $\gamma$ for cD+IC stars (solid line)
and galaxies (solid circles) for cluster C1. The same is shown for cluster
C2 by dashed line and solid triangles. An isotropic velocity distribution
has $\gamma$=1 indicated by a thin, dotted line.}  
\label{fig9}
\end{figure}
\begin{figure}
\epsfxsize=\columnwidth
\epsfbox{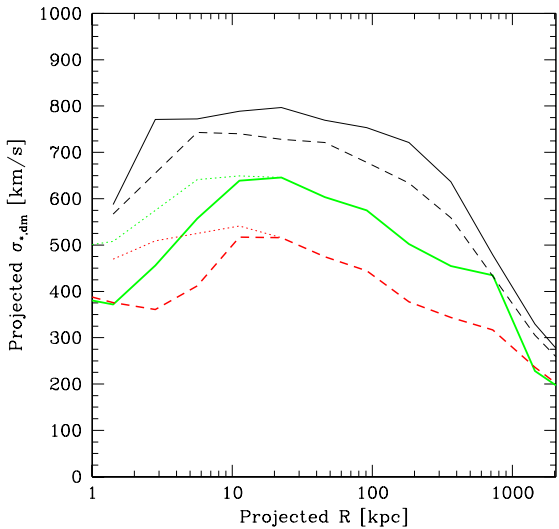}
\caption{Projected velocity dispersions for cluster C1 of cD+IC
stars (thick lines) and dark matter (thinner lines) along the minor axis 
(z-axis; dashed lines) and averaged along the x and y-axis (solid lines)
of the cD/cluster. Thin dotted lines mark the effect of making the 
correction for the excess young central stars. Statistical uncertainties
are $\Delta\sigma\la$20 km/s.}  
\label{fig10}
\end{figure}

The velocity distribution of the cD+IC stars in cluster C1 is highly
radially anisotropic at $r\ga$100 kpc, and, within the statistical 
uncertainties, this is
also the case for the cluster galaxies. We quantify this in Figure 9,
where we show 
$\gamma=\sqrt(2\sigma_r^2/(\sigma_{\phi}^2+\sigma_{\theta}^2))$,
where $\gamma^2$ is the ratio of the kinetic energy in radial and mean 
tangential (1D) motions,
respectively. For an isotropic velocity distribution $\gamma$=1. In
cluster C1 there is almost twice as much kinetic energy in the radial
direction than in the sum of the two tangential directions at $r\ga$100 kpc. 
Interestingly, within the statistical uncertainties
the same appears to hold for the galaxies, which at first seems surprising,
since one would expect that it is predominantly galaxies on radial orbits
which are tidally disrupted and transformed into cD+IC stars. The explanation
for this lack of difference in the orbital characteristics of the two
populations is likely the combined effect of radial infall of and dynamical
friction on cluster galaxies, but we defer a detailed discussion of this to
a forthcoming paper. We note that \cite{W.04} find the velocity distribution
of the IC stars to be more radially anisotropic than that of the galaxies in 
the outer parts of their simulated cluster.
In cluster C2 the velocity distribution of the cD+IC
stars is only mildly radially anisotropic and the velocity distribution of
the cluster galaxies is, within the statistical uncertainties, isotropic.
We also note that \cite{SL.97} found that the
velocity distribution of Galactic halo stars becomes tangentially
anisotropic in the outer halo --- the differences probably relate to
different accretion kinematics during the build-up of the various
haloes --- this will also be discussed in a forthcoming paper. 

Finally, concerning the flattening of the systems for cluster C1 
$\sigma_{\phi}$ is about 50\% larger than $\sigma_{\theta}$ for cD+IC stars
as well as galaxies, indicating that these systems are flattened by
anisotropic velocity distributions --- see below. For cluster C2 
$\sigma_{\phi}\simeq\sigma_{\theta}$, consistent with this cluster being
only mildly flattened. 

Observationally,
for the cD+IC stars one will only be able
to determine
line-of-sight velocities using planetary nebulae, not full 3D velocities.
For direct comparison with observations we show 
in Figure 10 the projected velocity dispersions of the cD+IC 
stars and (for comparison) of the dark matter in cluster C1 versus projected 
distance from
the center of the cD. By
dashed lines we show the azimuthally averaged velocity dispersions along
the minor axis of the cD. The projected stellar velocity
dispersion is 350-400 km/s at the center of the cD, increases to $\sim$500
km/s at $R\sim$10-30 kpc and then decreases gradually with increasing $R$
to about 300 km/s at $R_{\rm{vir}}$. An increase of the stellar velocity
dispersion with $R$ has been observed in some cDs, such as A2029 \citep{D79},
and has been interpreted as marking the transition from galactic to 
intracluster stars. The effect is less pronounced after correction for the
excess young central stars, but we note that these stars still contribute
to the gravitational potential and therefore boost the central velocity
dispersions. The projected velocity dispersion of the dark matter
follows a similar trend with $R$ as that of the stars, but is significantly
larger. As the stars and dark matter are moving in the same gravitational
potential this implies that the density distribution of dark matter is
significantly flatter than that of the cD+IC stars --- see below. 

The projected velocity dispersions along the major axis of the cD are shown 
by solid lines. They have been calculated as the average of the projected 
velocity dispersions along the x-axis and y-axis. Along the x-axis the 
velocity dispersions are calculated for stars and dark matter particles 
projecting onto $\pm$30 deg. wedges along the positive and negative y 
directions, and vice versa for dispersions along the y-axis. The major axis 
velocity dispersions follow similar trends as the minor axis ones.  
However, for the cD+IC stars it is significantly larger and for the dark
matter halo somewhat larger than along the minor axis. 
This shows that both systems
are flattened by anisotropic velocity dispersions and, as stated above,
oriented in similar ways. For cluster C2 the findings are similar (not
shown) though the cD is not aligned with the cluster in this case and
both systems are only mildly flattened.

Summarizing our findings for the Virgo-like cluster, the typical projected
velocity dispersion for the cD+IC stars at $R\ga$50 kpc is 300-600 km/s 
depending on orientation and projected distance from the cluster center.
\cite{F.00} find a (projected) velocity dispersion of 752$\pm$110 km/s
for 23 IC planetary nebulae in the Virgo-cluster at a projected distance
of about 150 kpc from M87. This value is somewhat larger than our prediction,
but it may be due to the ``real'' Virgo-cluster being kinematically
unrelaxed (see also section 3.3); the observational sample is too small to 
test for velocity substructure (Freeman 2004, private communication).

Since clusters are dynamically fairly young it is important to establish
that the results on kinematics etc. are robust, i.e. they do not depend on
the particular frame chosen for the analysis. To this end we analyzed
for each cluster two frames 1.5 and 3 Gyr before the present time, 
corresponding to redshifts 0.12 and 0.26, respectively. We find that none
of the results presented in this paper change in any significant way
when going to these earlier frames.
  
\begin{figure}
\epsfxsize=\columnwidth
\epsfbox{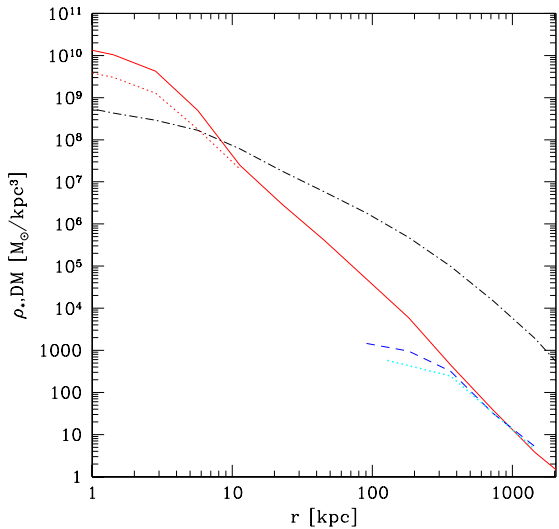}
\caption{Spherically averaged density profiles for cluster C1 of cD+IC
stars (solid line), dark matter (dot-dashed line), stars in cluster galaxies
(dashed line) and, with arbitrary normalization, {\it number} density of
cluster galaxies (dotted line). The result of applying the correction for the
excess, young central stars is shown by the thin, dotted line.}  
\label{fig11}
\end{figure}
\subsection{Density Distributions of the Intracluster Stars, Cluster Galaxies
and the Dark Matter}
In Figure 11 we show for cluster C1 the spherically averaged density
distributions of the cD+IC stars (with and without the correction for the 
excess, young central stars), the stars in cluster galaxies, the dark
matter and, with an arbitrary normalization, the {\it number} density of
cluster galaxies. As suggested previously, the number density profile of 
cluster galaxies in the inner parts of the cluster as well as the dark matter 
density profile in general are significantly
flatter than the density profile of cD+IC stars. For the cluster galaxies
this is a manifestation of the transformation of galaxies to cD+IC stars
by tidal disruption in the inner parts of the cluster. 

Moreover, it follows from Fig.11 that the average (mass) density of stars
in cluster galaxies only exceeds that of the IC stars beyond about half
the virial radius - this is in good agreement with the findings of
\cite{M.04}. 

\begin{figure}
\epsfxsize=\columnwidth
\epsfbox{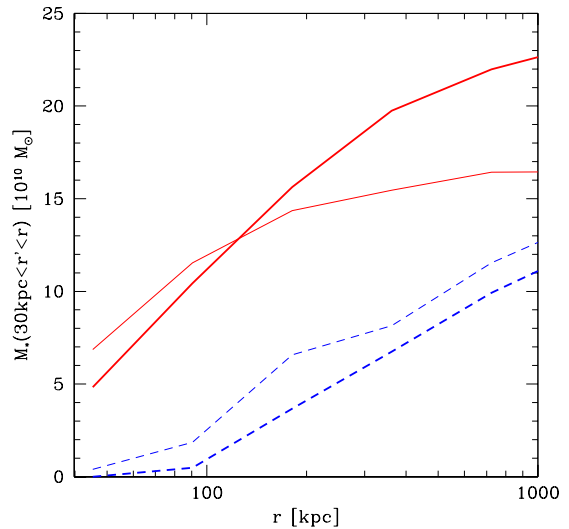}
\caption{For a $M_{\rm{vir}}$=3.4x10$^{13}$ $h^{-1}$M$_{\odot}$ (T=1.1 keV)
group is shown at $z$=0.1 the cumulative mass in cD+IC stars (solid lines) and 
in stars in cluster galaxies (dashed lines) outside of $r$=30 kpc for
the normal resolution (thin lines) and high resolution (thick lines)
simulations, respectively.}  
\label{fig12}
\end{figure}
\subsection{A Numerical Resolution Test}
It is important to test whether the properties of the IC star population
presented in this paper depend much on the numerical resolution
of the simulations. 
In the present paper, we address resolution effects by discussing a smaller
system, a group of virial mass {\mbox {$3.4 \times 10^{13} h^{-1} \Msun$}}
and emission--weighted temperature 1.1~keV. The group was run at the same
``standard'' resolution as clusters C1 and C2, as well as with eight times
higher mass and two times higher force resolution. This results in particle 
masses of $m_{\rm{gas}}$=$m_*$=3.1x10$^7$ and 
$m_{\rm{DM}}$=2.3x10$^8$ $h^{-1}$M$_{\odot}$ and gravity softening lengths 
of 1.4, 1.4 and 2.7 $h^{-1}$kpc, respectively. 
In order to enable an optimal comparison between the normal and high
resolution runs only Fourier modes up the Nyquist wavenumber of the
normal resolution simulation were used to prepare the initial conditions
for the high resolution run (i.e., additional high-wavenumber modes up to
the Nyquist wavenumber of the high resolution simulation were {\it not}
added to the Fourier modes). 

The resulting group looks like a ``scaled down'' version of clusters C1 and
C2. In particular, it has a prominent central galaxy, which we denote 
as the ``cD'', like in the cluster simulations --- in this respect it is
similar to the so-called ``fossil'' groups \citep[e.g.,][]{J.03}.

In Figure 12 we show the cumulative mass
of cD+IC stars and stars in galaxies outside of $r$=30 kpc for the normal
and high resolution runs at $z$=0.1 (at $z$=0 one large group galaxy is
merging with the cD). There is reasonable agreement between the runs ---
the somewhat larger mass in IC stars in the high resolution simulation
at $r\ga$200-300 kpc is likely due to the better resolution of star-forming
gas in the lower over-density regions which come to populate the outer
parts of the group. 

As one goes to higher resolution one might naively expect that the fraction
of IC stars in the simulations decreases as the cluster galaxies are
increasingly well resolved. The interesting indication from Figure 12 is,
however, that if anything the fraction of IC stars slightly {\it increases}
with increasing numerical resolution.

\begin{figure}
\epsfxsize=\columnwidth
\epsfbox{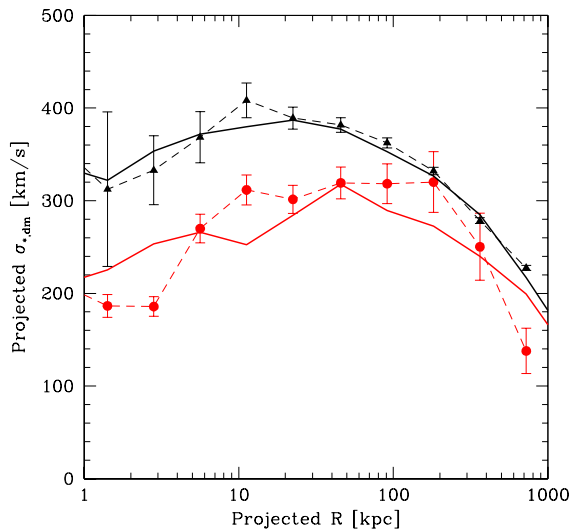}
\caption{Velocity dispersions projected along the minor axis of the ``cD''
for the cD+IC stars (lower curves and symbols) and dark matter (upper curves
and symbols) in the group. Results for the normal resolution simulation are 
connected by dashed lines and are shown as solid circles with errorbars
(stars) and solid triangles with errorbars (dark matter). Results
for the high resolution simulation are shown by solid lines.}  
\label{fig13}
\end{figure}
Shown in Figure 13 at $z$=0.1 are the azimuthally averaged velocity 
dispersions 
of cD+IC stars and dark matter projected along the minor axis of the cD for
the normal and high resolution simulations, and with the correction for
excess young central stars discussed previously applied. The agreement is 
overall reasonable, especially when it is taken into account that there
are at $z$=0.1 just 1235 cD+IC stars inside of the virial radius in the normal
resolution simulation of this group. This is a factor 10-20 less than the
corresponding numbers for clusters C1 and C2. The increase in central velocity
dispersion ($R\la$10 kpc) for the high resolution simulation, in particular
for the stars, is likely an effect of the increased force resolution.  
Note also the qualitative similarity between the velocity dispersion profiles
shown in Fig.~13, and the results for cluster C1 (Fig.~10).
  
Finally, as a test of resolution dependence of chemical evolution and 
metal abundances we show in Figure 14 the cumulative mass of Iron
in cD+IC stars and stars in galaxies outside of $r$=30 kpc for the normal
and high resolution runs. 
There is reasonable agreement between the
runs and we note that in relation to Figure 12 this is a nontrivial result,
since a significant fraction of the iron produced by the stars ends up
in the hot ICM (Paper I). 

In general, as mentioned at the beginning of section 3, we carry out
all the analysis of the (normal resolution) cluster simulations C1 and C2 
presented in this paper for the normal and high resolution
group simulations also. On this basis we find that all results given in 
this paper appear largely robust to resolution changes. Ultimately, however,
this can only be properly checked by running the ``production'' simulations
C1 and C2 at higher numerical resolution. To this end we have initiated a
high resolution simulation of C1 (with similar numerical characteristics
as the above high-resolution group simulation) --- at $z$=1.3 we find no significant
differences between the ``standard'' and high-resolution simulations of C1 in
relation to the results presented in this paper. 
\begin{figure}
\epsfxsize=\columnwidth
\epsfbox{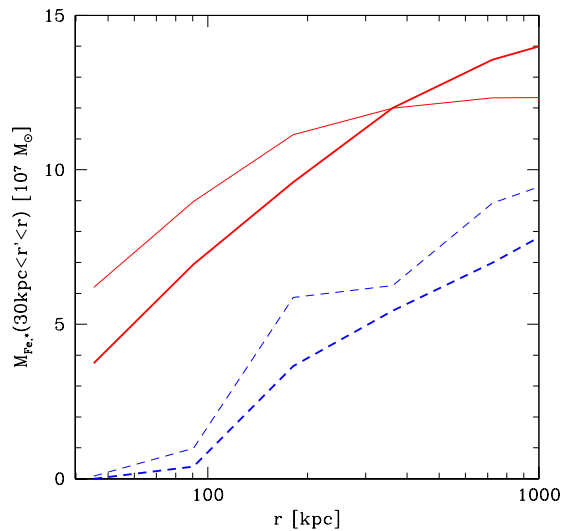}
\caption{Same as Figure 12, except that cumulative masses of Iron are shown.}  
\label{fig14}
\end{figure}
\section{Conclusions}
In this paper we have discussed the properties of the
intracluster (IC) stellar populations. Our results are based on cosmological
simulations of galaxy clusters including self-consistently metal-dependent
atomic radiative cooling, star-formation, supernova driven galactic 
super-winds, non-instantaneous chemical evolution and the effects of a
meta-galactic, redshift dependent UV field. In relation to modelling the
properties of the IC stars this is an important step forward with respect to 
previous theoretical works on the subject which 
were based on purely N-body (collisionless) simulations, apart from the
recent, interesting work by \cite{M.04}.

The main results from our simulations of a Virgo-like (C1) and a (sub)
Coma-like (C2) galaxy cluster regarding the IC stellar 
populations are as follows: 

The intracluster (IC) stars are found to contribute 20-40\% of the
total cluster B-band luminosity at $z$=0 and to form at a mean redshift
$\bar{z}_f\sim$3, somewhat larger than the mean formation redshift of the 
stars in cluster galaxies 
which is about 2.5. This difference corresponds
to a time span of about 0.5 Gyr. \cite{M.04} find somewhat lower mean 
formation redshifts
of 1.9 and 1.7, respectively --- the reason for this discrepancy between
the two works remains to be identified. 

We calculate UBVRIJHK surface brightness profiles of the IC star populations 
and find that the profile of the larger cluster matches the observed V-band
profile of Abell 1413 ($T\simeq$8 keV) quite well. For the Virgo-like
cluster we fail, however, to match the flat profile between 200 and 400
kpc projected cluster-centric distance observationally (but indirectly) 
inferred for the Virgo cluster by \cite{A.02} from planetary nebulae counts.
We note though that the galaxy distribution in Virgo is also unusually
flat \citep{B.87}.
The V-band surface brightness
is found to reach V$_{lim}\simeq$28.3 mag/arcsec$^2$ 
\citep{F.02} in cluster C2 at $r\sim$350 kpc and in cluster C1 at 
$r\sim$250 kpc. 
The typical colour of the IC stellar population is B-R=1.4-1.5,
comparable to the colour of sub-$L^*$ E and S0 galaxies.
The mean Iron abundance of the
IC stars is about solar in the central part of the cluster ($r\sim$100 kpc)
gently decreasing to about half solar at the virial radius. The IC stars
are $\alpha$-element enhanced with, e.g, [O/Fe] increasing slightly
with $r$ and characterized by a typical [O/Fe]$\sim$0.4 dex.

The IC stars are kinematically significantly colder than the cluster galaxies:
The velocity dispersions of the IC stars are in the inner parts of the
cluster ($r\sim$100-500 kpc) only about half of those of the cluster galaxies
increasing slightly to about 70\% at $r$=1-2 Mpc. The typical projected
velocity dispersion in the Virgo-like cluster at $R\ga$50 kpc is 300-600 km/s
depending
on orientation and projected distance from the cluster center. Rotation is
found to be dynamically insignificant for the IC stars. The velocity 
distributions of IC stars {\it and} clusters galaxies are in one cluster
highly radially anisotropic, in the other close to isotropic.

A test simulation of a $T\simeq$1.1 keV group at higher numerical resolution 
indicates that the results presented are largely robust to resolution changes.
Work is in progress to simulate also at higher resolution
the two clusters discussed in this paper. Moreover, we are in the process of
enlarging significantly our sample of galaxy clusters.  
\section*{Acknowledgements}
We have benefited considerably from discussions with Magda Arnaboldi, 
John Feldmeier, Ken Freeman, Anthony Gonzalez, Giuseppe Murante and Beth
Willman.
Moreover, the comments by the anonymous referee significantly improved the
presentation of our results. 

All computations reported in this paper were performed on the IBM SP4
facility provided by Danish Center for Scientific Computing (DCSC). We 
gratefully acknowledge the abundant access to computing time on this
system. This work was supported by Danmarks Grundforskningsfond through
its support for the establishment of the Theoretical Astrophysics Center
(TAC), and the Villum Kann Rasmussen Foundation.

\label{lastpage}


\begin{thebibliography}{99}
\bibitem[\protect\citeauthoryear{Arnaboldi}{2004}]{A04}
  Arnaboldi, M., 2004, IAU Symp., {\bf 217}, Recycling intergalactic and 
interstellar matter, eds. P.-A. Duc, J. Braine, and E. Brinks., p.54
\bibitem[\protect\citeauthoryear{Arnaboldi et al.}{2002}]{A.02}
  Arnaboldi, M. et al., 2002, AJ, 123, 760
\bibitem[\protect\citeauthoryear{Arnaboldi et al.}{2003}]{A.03}
  Arnaboldi, M. et al., 2003, AJ, 125, 514
\bibitem[\protect\citeauthoryear{Binggeli et al.}{1987}]{B.87}
  Binggeli, B., Tammann, G.A., \& Sandage, A., 1987, AJ, 94, 251
\bibitem[\protect\citeauthoryear{Binney \& Tremaine}{1987}]{BT87}
 Binney, J., \& Tremaine, S. 1987, Galactic Dynamics. Princeton Univ.
 Press, Princeton
\bibitem[\protect\citeauthoryear{Borgani et al.}{2004}]{B.04}
  Borgani, S., 2004, MNRAS, 348, 1078
\bibitem[\protect\citeauthoryear{Bryan et al.}{1998}]{BN98}
 Bryan G.L., Norman M.L., 1998, ApJ 495, 80
\bibitem[\protect\citeauthoryear{Cleary \& Monaghan}{1999}]{CM99}
  Cleary, P. W., \& Monaghan, J. J., 1999, Journal of Computational Physics,
  148, 227
\bibitem[\protect\citeauthoryear{Dressler}{1979}]{D79}
  Dressler, A., 1979, ApJ, 231, 659
\bibitem[\protect\citeauthoryear{Dubinski}{1998}]{D98}
  Dubinski, J., 1998, ApJ, 502, 141
\bibitem[\protect\citeauthoryear{Dubinski et al.}{2003}]{D.03}
  Dubinski, J., Koranyi, D., \& Geller, M., 2003, IAU symposium, 208, 237
\bibitem[\protect\citeauthoryear{Durrell et al.}{2002}]{D.02}
  Durrell, P.R., Ciardullo, R., Feldmeier, J.J., Jacoby, G.H., \& Sigurdsson, 
S., 2002, ApJ, 570, 119
\bibitem[\protect\citeauthoryear{Ettori \& Fabian}{2000}]{EF00}
  Ettori, S., \& Fabian, A.C., 2000, MNRAS, 317, 57
\bibitem[\protect\citeauthoryear{Feldmeier et al.}{2002}]{F.02}
  Feldmeier, J.J., et al., 2002, ApJ, 575, 779
\bibitem[\protect\citeauthoryear{Feldmeier et al.}{2004a}]{F.04a}
  Feldmeier, J.J., et al., 2004a, ApJ, 609, 617
\bibitem[\protect\citeauthoryear{Feldmeier et al.}{2004b}]{F.04b}
  Feldmeier, J.J., Ciardullo, R., Jacoby, G.H., \& Durell, P.R., 2004b, ApJ, 
in press (astro-ph/0407274)
\bibitem[\protect\citeauthoryear{Ferguson et al.}{2002}]{Fer.02}
  Ferguson, A.M.N., Irwin, M.J., Ibata, R.A., Lewis, G.F., \& Tanvir, N.R., 
  2002, AJ, 124, 1452
\bibitem[\protect\citeauthoryear{Freeman et al.}{2000}]{F.00}
  Freeman, K.C., et al., 2000, ASP Conf. Series, 197, 389
\bibitem[\protect\citeauthoryear{Gal-Yam et al.}{2003}]{G.03}
  Gal-Yam, A., et al., 2003, AJ, 125, 1087
\bibitem[\protect\citeauthoryear{Garilli et al.}{1997}]{G.97}
  Garilli, B., et al., 1997, AJ 113, 1973
\bibitem[\protect\citeauthoryear{Girardi et al.}{2002}]{G.02}
  Girardi, L., et al., 2002, A\&A, 391, 191
\bibitem[\protect\citeauthoryear{Gladders et al.}{1998}]{G.98}
  Gladders, M.D., et al., 1998, ApJ, 501, 571
\bibitem[\protect\citeauthoryear{Gonzalez et al.}{2000}]{G.00}
  Gonzalez, A.H, et al., 2000, ApJ, 536, 561
\bibitem[\protect\citeauthoryear{Gonzalez, Zabludoff \& Zaritsky}{2004}]{G.04}
  Gonzalez, A.H, Zabludoff, A.I., \& Zaritsky, D. 2004, ApJ, submitted 
(astro-ph/0406244)
\bibitem[\protect\citeauthoryear{Haardt \& Madau}{1996}]{HM96} 
  Haardt, F., \& Madau, P. 1996, ApJ, 461, 20
\bibitem[\protect\citeauthoryear{Helmi et al.}{1999}]{H.99}
  Helmi, A., White, S.D.M., de Zeeuw, P.T., \& Zhao, H., 1999, Nature, 402, 53
\bibitem[\protect\citeauthoryear{Helmi et al.}{2003}]{H.03}
  Helmi, A., White, S.D.M., \& Springel, V., 2003, MNRAS, 339, 834
\bibitem[\protect\citeauthoryear{Jones et al.}{2003}]{J.03}
  Jones, L.R., Ponman, T.J., Horton, A., Babul, A., Ebeling, H. \& 
Burke, D.J., 2003, MNRAS, 343, 627
\bibitem[\protect\citeauthoryear{Kormendy}{1977}]{K77}
  Kormendy, J., 1977, ApJ, 218, 333
\bibitem[\protect\citeauthoryear{Lia, Portinari \& Carraro}{2002a}]{L.02} 
 Lia, C., Portinari, L., \& Carraro, G.  2002a, MNRAS, 330, 821
\bibitem[\protect\citeauthoryear{Lia, Portinari \& Carraro}{2002b}]{LPCerr} 
 Lia, C., Portinari, L., \& Carraro, G.  2002b, MNRAS, 335, 864
\bibitem[\protect\citeauthoryear{Lin \& Mohr}{2004}]{LM2004} 
 Lin Y.-T., Mohr J.J., 2004, ApJ in press (astro-ph/0408557)
\bibitem[\protect\citeauthoryear{Mackie}{1992}]{M92} 
Mackie, G. 1992, ApJ 400, 65
\bibitem[\protect\citeauthoryear{McNamara}{2004}]{Mc04} 
 McNamara, B. R. 2004, in ``The Riddle of Cooling Flows in Galaxies and
Clusters of Galaxies'', Charlottesville, Virginia, USA, Eds. T.H.Ruprecht,
J.C.Kempner \& N.Soker
\bibitem[\protect\citeauthoryear{Mori et al.}{1997}]{M.97} 
Mori, M., Yoshii, Y., Tsujimoto, T., \& Nomoto, K.  1997, ApJ, 478, L21
\bibitem[\protect\citeauthoryear{Murante et al.}{2004}]{M.04} 
 Murante, G., et al.,  2004, ApJ, 607, L83
\bibitem[\protect\citeauthoryear{Napolitano et al.}{2003}]{N.03} 
 Napolitano, N.R., et al., 2003, ApJ, 594, 172
\bibitem[\protect\citeauthoryear{Oemler}{1976}]{O76}
  Oemler, A. Jr., 1976, ApJ, 209, 693
\bibitem[\protect\citeauthoryear{Romeo, Portinari \& Sommer-Larsen}{2004}]
{RPSL04} 
 Romeo, A.D., Portinari, L., \& Sommer-Larsen, J., 2004, 
 MNRAS, submitted (astro-ph/0404445, Paper II)
\bibitem[\protect\citeauthoryear{Romeo et al.}{2004}]{R.04} 
 Romeo, A.D., Sommer-Larsen, J., Portinari, L., \& Antonuccio, V., 2004, 
 MNRAS, in preparation (Paper I).
\bibitem[\protect\citeauthoryear{Sommer-Larsen \& Dolgov}{2001}]{SLD01} 
 Sommer-Larsen J., Dolgov A., 2001, ApJ 551, 608
\bibitem[\protect\citeauthoryear{Sommer-Larsen et al.}{1997}]{SL.97} 
 Sommer-Larsen, J., Beers, T.C., Flynn, C., Wilhelm, R.,
\& Christensen, P. R. 1997, ApJ, 481, 775 
\bibitem[\protect\citeauthoryear{Sommer-Larsen, G\"{o}tz \& Portinari}{2003}]
{SL.03}
  Sommer-Larsen J., G\"{o}tz M., Portinari L., 2003, ApJ, 596, 46
\bibitem[\protect\citeauthoryear{Springel \& Hernquist}{2002}]{SH02} 
 Springel, V., \& Hernquist, L.,  2002, MNRAS, 333, 649
\bibitem[\protect\citeauthoryear{Tornatore et al.}{2004}]{T.04}
  Tornatore, L., Borgani, S., Matteucci, F., Recchi, S., \& Tozzi, P., 2004, 
  MNRAS, 349, L19
\bibitem[\protect\citeauthoryear{Valdarnini}{2003}]{V03}
  Valdarnini, R., 2003, MNRAS, 339, 1117
\bibitem[\protect\citeauthoryear{Willman et al.}{2004}]{W.04}
  Willman, B., Governato, F., Wadsley, J., \& Quinn, T., 2004, 
  MNRAS, submitted (astro-ph/0405094)
\bibitem[\protect\citeauthoryear{Zaritsky, Gonzalez \& Zabludoff}{2004}]{Z.04}
  Zaritsky, D., Gonzalez, A.H, \& Zabludoff, A.I., 2004, ApJ, 613, L93 
\end{thebibliography}
\end{document}